\documentstyle[prl,aps,twocolumn,floats]{revtex}
\input psfig.tex
\begin{document}
\preprint{PUPT-94-1456}
\twocolumn[\hsize\textwidth\columnwidth\hsize\csname @twocolumnfalse\endcsname
\title{Z Condensation \\ and \\ Standard Model Baryogenesis}
\author{Salim Nasser and Neil Turok}
\address{Joseph Henry Laboratories, Princeton University, Princeton, NJ 08544}
\date{\today}
\maketitle

\begin{abstract}
The standard model satisfies Sakharov's conditions for baryogenesis
but the CP violation in the KM matrix appears too small to account
for the observed asymmetry. In this letter we explore a mechanism
through which CP violation can be greatly amplified.
 First CP is {\it spontaneously}
broken through dynamical effects on bubble walls, and
the two CP conjugate phases grow through phase ordering.
Direct competition between macroscopic regions of both phases
then amplifies
the microscopic CP violation to a point where one of the
two phases predominates. This letter is devoted to a demonstration
that spontaneous CP violation may indeed occur on
propagating bubble walls via the formation of a condensate
of the longitudinal $Z$ boson.
\\
\end{abstract}
]

A fascinating development in particle theory
over the last few years has been the
realisation that the
standard model,  possibly with a minimal
extension of the Higgs sector, has the potential to produce the observed
matter-antimatter asymmetry in the universe.
At the heart of the mechanism is the remarkable anomaly
structure of the Weinberg-Salam theory, which ties changes in the
topology of the gauge and Higgs fields
 to changes in the baryon number of
the universe
\cite{'t Hooft}
If these baryon number violating processes were somehow
 biased at the electroweak phase transition, and subsequently
 turned off, an  asymmetry would be generated
 rather naturally.

Interest in this idea grew when explicit mechanisms
producing the required bias were proposed, in minimal extensions of
the standard model involving extra Higgs fields
(for reviews see
\cite{Turok,Cohen}).
Two simple arguments seem to exclude the minimal standard model.
The $CP$ violation due to the Kobayashi-Maskawa
matrix is very small, suppressed by many powers of light quark
masses and mixing angles. A naive estimate assuming analyticity
in fermion masses indicates that
CP violation should enter in the combination

\begin{equation}
d_{CP} = J \Pi_{i\neq j} (m^2_i-m^2_j) \Lambda^{-12}
\end{equation}

where the product is over like charged quarks,
and $J \sim 10^{-5}$ is a product of $KM$ angles. Taking the
scale $\Lambda$ to be the temperature $T \sim 100$ GeV,
 one finds $d_{CP} \sim
10^{-19}$ leading to a baryon symmetry well below the
required value $\sim 10^{-10}$.
Second, the requirement that the
anomalous  processes be suppressed after the phase transition
imposes an upper bound on the mass of the Higgs boson - estimated
from the one loop potential
\cite{Dine}
to be $M_H < 35$ GeV, in conflict
with the LEP bound $M_H>60$ GeV. But
recent lattice studies
\cite{Farakos}
indicate that the electroweak transition is far
more strongly first order,
relaxing the upper bound on $M_H$, perhaps up to 100 GeV.

The
prospect of calculating the baryon asymmetry in
terms of known experimental parameters make the search for
a mechanism which amplifies (1) attractive.
Recently, Farrar and Shaposhnikov \cite{Farrar}  pointed out
one very interesting mechanism involving
very low momentum quasiparticles in the
quark plasma.
In their calculation, the scale $\Lambda$ is
provided by a light quark mass, enormously amplifying $d_{CP}$.
However
it appears that when
damping  due to strong scattering is included
a negligible asymmetry results \cite{Gavela,Huet}.

One might
conclude that
generating the observed asymmetry in the standard model is
hopeless. We believe this is premature, and in this letter
propose a new mechanism through
which $CP$ violation in the $KM$ matrix
may be greatly amplified in the standard model
(or extensions such as the minimal supersymmetric version)
 through
macroscopic physical effects qualitatively different than those
previously considered.

There is another very {\it large} number involved in the
electroweak transition,
namely the horizon $R_H$ in
units of $T^{-1}$.
The electroweak phase transition proceeds via bubble nucleation
and growth. Calculations based on the
perturbative potential show that
bubbles fill space
 when their typical radius
is $\sim 10^{-4} -10^{-5}R_H$.
 If the phase transition
is more strongly first order, this could increase to a value
closer to  $R_H
\sim 10^{16} T^{-1}$.

\begin{figure}[htbp]
\centerline{\psfig{file=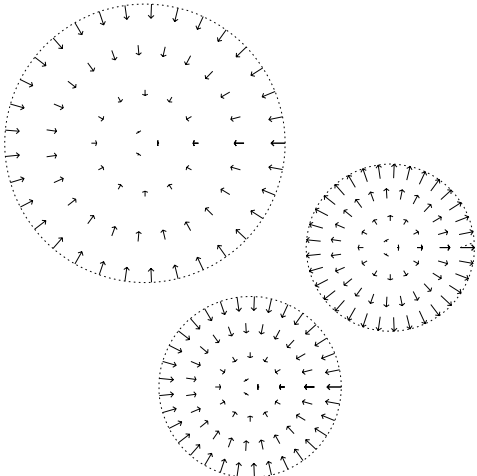,width=3.5in}}
\caption{Bubbles of the broken symmetry phase growing at the electroweak
transition, with a condensate of longitudinal $Z$ bosons on their
surfaces. When large bubbles collide, `in' and `out' phases are
placed in contact, leading to competition between macrocopic
phases, which greatly amplifies $CP$ violating effects.}
\label{fig:f1}
\end{figure}

Our scenario makes use of this as follows.
As bubbles form and grow, collisions with fermions in the plasma
lead to a $CP$ violating instability on the bubble wall, namely
 a condensate of longitudinal  $Z$ bosons. The $Z$ field can
 point either `out' or `in' (Figure \ref{fig:f1}). Patches of each phase
 initially cover each bubble, separated by one dimensional phase
 boundaries, which are in fact $Z$-magnetic flux tubes. As the
bubble expands, small patches shrink away under the tension
of the phase boundary, while larger patches grow with the
bubble. The long time available before bubbles collide allows
each bubble to be covered by {\it macroscopic} regions of the `in' and the
`out' phases. Even if each bubble becomes completely ordered,
macroscopic regions of both phases are placed in competition
when large bubbles collide as the transition completes.

Now as a given phase moves through the plasma, its terminal
velocity is given by $v_t=\frac{\Delta P}{\Gamma}$ where $\Delta
P$ is the pressure difference between the false and true vacua
and where the
frictional drag/unit area is given by $F_{\rm drag}=\Gamma v$.  As a result
of the $CP$ violation in the $KM$ matrix, we expect the `in' and
`out' phases to feel slightly different $\Gamma$'s,
$\frac{\Gamma_+ -\Gamma_-}{\Gamma_+ + \Gamma_-} \sim d_{CP}$.
It follows that the two phases move through the plasma with
slightly different velocities $|v_+ - v_-| \sim d_{CP}$.  Now
consider the neighbourhood of a phase boundary.  When the bubbles
are large enough, we may take them to be planar as far as the
local dynamics of the phase boundary go. As the bubbles grow, the
`out' phase begins to bulge over the `in' phase,
the height of the bulge being given by $\sim d_{CP}
 t$. After a time of order one expansion time,
this becomes comparable to the scale which
enters in determining the shape of the bubble wall,
$l=\frac{\sigma}{\Delta P}$, where $\sigma$ is the surface
tension.
The bulge then gives rise
to a net tangential force on the phase
boundary, causing it to sweep across the bubble at a velocity of
the order of the bubble wall velocity.
Thus the faster moving phase
will (again in a time of the order of an expansion time),
completely overtake the slower moving phase \cite{Foot1}.
In this scenario,
as the electroweak phase transition nears completion,
bubble surfaces of one $CP$ violating variety predominate.
Notice that the only place
the small $CP$-violating parameter comes in is in ensuring that
the `out' and `in' phases expand at slightly different rates.

The longitudinal $Z$ condensate leads straightforwardly
 to the production of a baryon asymmetry
through both `local'
\cite{Zadrozny,Mclerran}
and `nonlocal' mechanisms
\cite{Kaplan,Joyce}, because it
  couples to fermions in exactly the same way
that the $CP$ violating phase $\theta$ does in two-Higgs scenarios.

The remainder of this letter will be devoted to a demonstration
that longitudinal $Z$
condensate  can form on propagating bubble walls.
The longitudinal $Z$ may be defined
by the formula $i {\varphi }^* \overline{\cal D}_\mu {\varphi} \equiv
- 2 g_A \phi^2 Z_\mu$, where $\phi= \sqrt{2 \varphi^\dagger \varphi}$
and $g_A =  {1\over 4}
(g_1^2+g_2^2)^{1\over 2}$, with
$g_1, g_2$ the $U(1)_Y $ and $SU(2)_W$ gauge couplings.
We now consider the classical field equation
for $Z_\mu$, in background of a bubble wall
propagating through the plasma.
In the wall rest frame we  take the Higgs field
to have a fixed profile $\phi(z)$, its width $L_w$
being of order the inverse
of the Higgs mass $m_H$ in the broken phase. The wall is treated
as planar, which is a good approximation once a bubble gets large.
We consider a  $Z$ field $Z^\mu =
(0,0,0,Z(z))$. We may consistently
drop the $W^{\pm}$ boson and
photon fields, because the fermion currents
induced by the $Z$ are neutral and so only source the $Z$. Likewise
the $Z$ makes no contribution to the $SU(2)$ currents in the
absence of a $W$ field.

Next, we drop spatial derivative terms, which are suppressed
by powers of $m_H/m_Z$, and $m_H/m_{top}$ relative to
the $Z$ mass term and the top quark currents which dominate.
The equation of motion then becomes

\begin{equation}
\ddot{Z} = J_Z(Z) +J_F(Z)
\label{eqom}
\end{equation}

where $J_Z(Z)
=-4 g_A^2 \phi^2  Z $
is the current carried by the $Z$ condensate and
$J_F(Z)
= g_A \overline{\psi} \gamma^3 \gamma^5 \psi +g_V
\overline{\psi} \gamma^3 \psi$
is that carried by the
fermions.

In the absence of the $W^{\pm}$ bosons, CP invariance holds,
and so
$Z \rightarrow -Z$ is a symmetry \cite{Foot2}.
 Thus
there is a solution
$Z=J_F=0$. But we shall show that this is
unstable, so that a nonzero
$Z$ current develops. We expect this instability to
lead to one of  two stable solutions, related by
$Z \rightarrow -Z$, in which $J_F(Z)$ and
$J_Z(Z)$ are equal and opposite so that
the net current is zero.
The condition for an instability is simply that
$ m_{eff}^2  = 4 g_A^2 \phi^2 - (\partial J_F /\partial Z)|_{Z=0}<0$.

The fermion current $J_F$
is calculated by solving the time independent
Dirac equation in the presence of $\phi$ and $Z$ condensates.
We have used the WKB approximation as a
guide to the
behaviour expected, and checked it against results obtained
with an exact linear response function.

\begin{figure}[htbp]
\centerline{\psfig{file=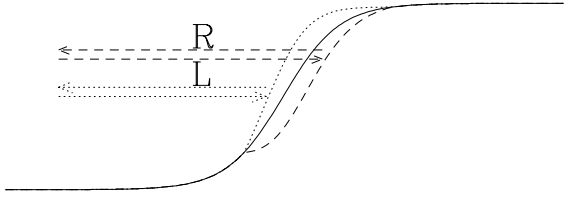,width=3.5 in}}
\caption{Profile of a bubble wall, showing the barriers seen by
particles with spins $S^z$ in the positive and negative $z$ directions.
Positive spin particles incident from the left get further
than negative spin particles, causing a positive chiral current
which acts to enhance the $Z$ condensate. This is the
origin of the instability.}
\label{fig:f2}
\end{figure}

The relevant Dirac equation is

\begin{equation}
(i \gamma^\mu (\partial_\mu + i Z_\mu(g_V +g_A\gamma^5)) -m(z)) \psi =0
\label{dirac}
\end{equation}

where $m(z) = y \phi(z)$ with $y$ the Yukawa coupling
We ignore for the moment
the one loop
thermal contributions to the fermion self energies.
{}From (\ref{dirac}) the following dispersion relation is found

\begin{equation}
\omega_\pm = \sqrt{ p_\perp^2 +(\sqrt{(p_z-g_V Z)^2 +m^2(z)} \mp g_A Z(z))^2}
\label{disp}
\end{equation}

for the two eigenstates, which have
spin $S^z =\pm {1\over 2}$  in a frame where $p_\perp=0$.
In the rest frame of the wall, both the energy $\omega$ and
the transverse momentum $p_\perp$ are conserved, and the
local value of the momentum $p_z$ varies according to
(\ref{disp}).
It is related to $p_{z,-\infty}$, the value of the momentum at $z=-\infty$,
by the formula $E= \sqrt{p_\perp^2 +p_{z,-\infty}^2}$.
Regions
for which $ p_{z,-\infty}< m(z) \mp g_A Z $ are classically
disallowed:  the $S^z=\pm {1\over2}$
excitations see a `barrier' $m(z) \mp g_A Z(z)$
respectively (Figure \ref{fig:f2}).

 The fact that
left and right handed fermions carry different charges now comes into
play - particles of given $S^z$ carry the same
current when travelling in either direction!
One can anticipate the destabilising
effect
- for positive $Z$ there is a
 region where only $S^z>0$ particles penetrate
from the left, creating a positive chiral current. This acts
through the equation of motion (\ref{eqom}) to further destabilise
the $Z$. Including  currents from antiparticles,
which are given by the substitutions
$J \rightarrow -J$; $g_{A,V} \rightarrow -g_{A,V}$,
multiplies the axial current by two, while cancelling the
vector current. The parity violation in the
$Z$-fermion coupling is crucial -
the charged vector bosons $W^{\pm}$ do not
destabilise the $Z$ because the
$Z/W^{\pm}$ equations of motion are parity invariant so that
left and right moving $W^{\pm}$ modes carry opposite currents.

A more detailed analysis reveals a competing stabilising effect
which exactly cancels the leading destabilising effect at zero
wall velocity.
The current carried per mode is given by
the classical formula  $J_z=- (\partial \omega /\partial Z)$
\cite{Foot3}.
Integrating over momenta this is (for $S^z>0$)

\begin{eqnarray}
J_F &=& -\int {d^3p\over (2 \pi)^3}  f(p) {\partial \omega_+ \over \partial Z}
\\
 &=& g_A \int {d^3p_{-\infty} \over (2 \pi)^3}
 \bigl({dp_z \over dp_{z,-\infty}} \bigr) f(p_{-\infty})
 {p_{z,-\infty} \over \sqrt{p_\perp^2 +p_{z,-\infty}^2} }
\label{current}
\end{eqnarray}

(the phase space density $f$ is constant
along particle trajectories). The only $Z$ dependence
is now in the
 Jacobian, which is just the ratio of the group
 velocity at infinity to that locally,

\begin{equation}
v_z={\partial \omega \over
\partial p_z} = {\sqrt{(p_{z,-\infty} +g_A Z)^2 -m^2} \over
p_{z,-\infty} +g_A Z} \bigl({p_{z,-\infty} \over w}\bigr)
\label{gvel}
\end{equation}

The Jacobian represents
the enhancement of the local particle density due to a
`slowing down' effect.
 For positive $Z$, $S^z>0$  particles see a `well' and
 speed up, decreasing the chiral current, and $S^z<0$ particles
 see  a `barrier', slowing down and adding a negative chiral
 current.
Thus particles passing `over' the barrier (in either direction)
act to stabilise the $Z$ condensate, whereas particles
`bouncing' off the barrier tend to destabilise it.

Now we compute the net chiral current
in the WKB approximation.
 We assume the  barrier is monotonic ($|(dZ/dm)|<1$),
 and sum over particles and antiparticles of both spins,
 incident from both sides of the wall with thermal distributions
 at $z=\pm \infty$.
As already mentioned,
the leading effect, a square root divergence, cancels at zero
$v_w$ - there is a factor of two in the `bouncing'
contribution, cancelled by a minus one from particles going over the
barrier in each direction. But at finite $v_w$ more particles are encountered
from the left, enhancing the number of `bouncing' particles.
The dominant term occurs as $m(z)$ approaches $m_\infty$, and
to lowest order in $v_w$ is given by

\begin{equation}
J_F = {2 g_A N_C \over (2 \pi )^2} {m_\infty^2 v_w\over
e^{\beta m_\infty} +1} \sqrt{(m_\infty +g_A Z)^2 -m(z)^2 }
 -(Z\rightarrow -Z)
\label{jcur}
\end{equation}

where we sum over colors $N_C$.
Note that

\begin{equation}
{\partial J_F \over \partial Z}|_{Z=0}
\propto g_A^2 m_\infty^3 (m_\infty^2 -m(z)^2)^{-{1\over 2}} \propto
g_A^2 m_\infty^2 e^{z/2L}
\label{jcurb}
\end{equation}
diverging as $z\rightarrow \infty$, so that (within the WKB,
free particle approximation we have made)
the fermionic contribution to the effective $Z$ squared mass diverges,
due to the `bouncing' effect, and an instability always develops
sufficiently far behind the wall.

Several comments
are in order.
The destabilising term
is proportional to $m_\infty^2$, so the
top quark dominates.
We ignored the decay process
$t\rightarrow W+b$, which occurs once the $t$ gets a large
enough mass
on the wall.
The rate for this process
is ${1\over 16} \alpha_2 m_t^3/m_W^2(1-m_W^2/m_t^2)^2(1+2m_W^2/m_t^2)
\sim m_t/110$, and the timescale is long compared to
the time the top quarks spend on the wall, in the regime of
interest.
We
have treated the fermions bouncing
off the wall as free particles, ignoring
their interactions with the plasma. More
detailed calculations are needed to  reveal whether these interactions
strengthen or suppress the destabilising term.
One estimate is to
assume the relevant top quarks are emitted with a thermal
distribution a distance $D$ away from the point $z$. Then
 in (\ref{jcurb})
the exponential is replaced by $e^{D/2L_w}$, with
$D$ a rather short diffusion length,
smaller than $L_w$
(we need
the exponent to be $\sim 3$ for instability).
But this is too pessimistic - it ignores
the `pile-up' of tops in front of the wall,
particularly $p_z$ below the barrier.
In any case, a full semiclassical calculation including strong scattering in
the Boltzmann equation appears quite feasible.

We emphasise that unlike
the calculations of \cite{Farrar},\cite{Gavela} and \cite{Huet} which
are one dimensional, our calculations
are three dimensional, and phase space is dominated by
{\it large} perpendicular
momenta $p_\perp \sim 2T$.
At large $p$ the quark damping rate
\cite{Pisarski} is smaller, and
in the broken phase
the top's large mass makes it less sensitive to collisions at
low momentum transfer.
Note also that a difference from Farrar and Shaposhnikov
is that
the spontaneous $CP$ violation here is essentially a
{\it classical} effect, and does not rely on the
quantum mechanical coherence.
We expect it to be
less susceptible to destruction by strong scattering.

The WKB calculation is indicative of an interesting effect,
but since the instability
occurs near (and because of!) a classical turning point,
it is important to check it in a full quantum mechanical calculation.
 We used
an exact solution to the
Dirac equation in the background $\phi(z)^2 \propto
(1+e^{-a z})^{-1}$ in hypergeometric
functions, in order to compute the full linear response kernel
$K(x,y) \sim (\delta J_F(x) /\delta Z(y))$ \cite{Nasser},
 which enters the $Z$ equation as
\begin{equation}
\ddot{Z}(z) = g_A^2\bigl(- 4 \phi^2  Z(z) +
{T^2 N_C \over 4 \pi^2} \int dy K(z,y) Z(y)\bigr)
\label{jcurk}
\end{equation}

The full quantum problem is  nonlocal, and
the condition for instability now reads

\begin{equation}
{\int \int dx dy Z(x) K(x,y) Z(y) \over \int dx Z^2(x) (\phi^2(x)/
\phi^2_\infty)}
> {16 \pi^2 \phi_\infty^2 \over N_c T^2} \sim 100
\label{jcurs}
\end{equation}

for some trial function $Z(x)$, where we used
the bound \cite{Turok} $\phi_\infty \sim T/g_2 \sim 1.5 T$
at the transition - if $\phi_\infty$ were smaller, baryons would not
survive in the broken phase.

We have computed the integrals in (\ref{jcurs}) numerically,
using Mathematica to evaluate the kernel.
In order to focus
on the `bouncing' effect, we consider
the case where the only particles incident are those below the barrier
from the left, and we include both spins, particles and antiparticles.
We integrate over $0< p_\perp< \infty$
with a thermal distribution function and then numerically over
$p_{z,-\infty}$,
ignoring a phase space factor $x$Log$(1+e^{-x})$, with $x=p_{z,-\infty}/T$
which is of order unity, and in any case
could be enhanced by
a chemical potential.  We then compare the left hand side
of (\ref{jcurs}) with the leading term in
the WKB approximation for exactly the same quantity. (In
WKB, the neglect of particles passing over the barrier has the
effect of doubling the destabilising term).

In units where $T=1$,
we used $a=.1$ (wall width $L_w=10$), and $m_\infty=1$ for the
top mass in the broken phase. With  a Gaussian  trial function
exp$(-(x-x_0)^2/w^2)$, and $x_0=6 L_W$,
the left hand side of
(\ref{jcurs}) is greatest for $w \approx 2 L_W$, and has a value
179, compared to the WKB value of 161.
The quantum problem thus shows a destabilising
term even greater than the leading WKB approximation.
A comprehensive set of results for different
wall thicknesses will be presented in \cite{Nasser}.

Several outstanding questions remain for future work.
Exactly how does the $CP$ violation from the $KM$ matrix enter,
and distinguish the `in' and `out' phases of the
$Z$ condensate? In a one dimensional treatment (which likely
overestimates the effect), one can see from the calculations
of \cite{Huet} that a $Z$ independent term $\sim d_{CP}$ enters
in the equation of motion (\ref{eqom}), and we expect to find similar
$CP$ violating terms at other even powers of $Z$. We are investigating
the analogous three dimensional calculation.
Finally,  before the
baryon asymmetry may be accurately calculated, we
need to find the stable solutions for the $Z$ condensate.

\begin{acknowledgements}
We thank M. Bucher, D. Gross, I. Kogan, M. Joyce and
T. Prokopec for discussions.
This work  was partially  supported by
NSF contract PHY90-21984, and the David and Lucile Packard
Foundation.
\end{acknowledgements}


\begin{thebibliography}{99}
\bibitem{'t Hooft} G. 't Hooft. Phys. Rev. Lett., {\bf 37}, 8,
1976; A.D. Linde. Phys. Lett. {\bf 70B}, 306, 1977; V. Kuzmin, V.
Rubakov and M. Shaposhnikov, Phys. Lett. {\bf 155B}, 36 (1985);
F. Klinkhamer and N. Manton, Phys. rev {\bf D30}, 2212 (1984); P.
Arnold and L. Mclerran, Phys. Rev {\bf D37}, 1020.
\bibitem{Turok} N. Turok, in {\it Perspectives in Higgs Physics},
ed. G. Kane, pub. World Scientific, p. 300 (1992).
\bibitem{Cohen} A. Cohen, D. Kaplan and A. Nelson, Ann. Rev.
Nucl. Part. Sci. {\bf 43}, 27 (1993).
\bibitem{Dine} M. Dine, R. Leigh, P. Huet, A. Linde, and D.
Linde, Phys. Rev. {\bf D46}, 550 (1992); B-H. Liu, L. Mclerran
and N. Turok, Phys. Rev {\bf D46}, 2668 (1992).
\bibitem{Farakos} K. Farakos, K. Kajantie, K. Rummukainen and M.
Shaposhnikov, Nucl. Phys. {\bf B407}, 356 (1993); preprint
CERN-TH.7244/94.
\bibitem{Farrar} G. Farrar and M. Shaposhnikov, Phys. Rev. Lett.
{\bf 70}, 2833 (1993); {\bf 71}, 210(E) (1993); preprint
CERN-TH.6732/93.
\bibitem{Gavela} M.B. Gavela, P. Hernandez, J. Orloff and O.
Pene, preprint CERN-TH.7081/93.
\bibitem{Huet} P. Huet and E. Sather, preprint SLAC-PUB-6179
(1994).
\bibitem{Zadrozny} N. Turok and J. Zadrozny, Phys. Rev. Lett.
{\bf 65}, 2331 (1990).
\bibitem{Mclerran} L. Mclerran, M. Shaposhnikov, N. Turok and M.
Voloshin, Phys. Lett. {\bf 256B}, 451 (1991).
\bibitem{Kaplan} A. Cohen, D. Kaplan and A. Nelson, Nucl. Phys.
{\bf B373}, 453 (1992).
\bibitem{Joyce} M. Joyce, T. Prokopec and N. Turok, Princeton
preprint PUP-TH-1437 (1994).
\bibitem{Pisarski} R. Pisarski, Phys. Rev. {\bf D47}, 5589 (1992).
\bibitem{Nasser} S. Nasser and N. Turok, in preparation, 1994.
\bibitem{Foot1} We have
confirmed these expectations in a calculation which will
be described in \cite{Nasser}.
\bibitem{Foot2} Consider a spherical bubble
as in (Figure\ref{fig:f1}): P leaves the $Z$ configuration invariant,
 but C reverses it.
\bibitem{Foot3} In a box where the $Z$ field is turned on adiabatically,
$Z=Z(t)$, the canonical momentum of a mode doesn't change, but
the work done is $-\int dt J_z {\cal E}_z$, with ${\cal E}$ the
electric field. Using ${\cal E} = -\partial_t Z$ gives the wanted result.
\end{thebibliography}
\end{document}